\documentclass[fleqn,usenatbib]{mnras}
\usepackage{newtxtext,newtxmath}
\usepackage[T1]{fontenc}
\DeclareRobustCommand{\VAN}[3]{#2}
\let\VANthebibliography\thebibliography
\def\thebibliography{\DeclareRobustCommand{\VAN}[3]{##3}\VANthebibliography}

\usepackage{graphicx}	
\usepackage{amsmath}	
\usepackage{subcaption}
\title{Conducting High Frequency Radio SETI Searches using ALMA}

\author[L. Mason et al.]{Louisa A. Mason$^{1}$, Michael A. Garrett,$^{1,2}$, Kelvin Wandia $^{1}$ and Andrew P. V. Siemion $^{1,3,4,5,6}$
\\
$^{1}$Jodrell Bank Centre for Astrophysics, Department of Physics and Astronomy, Alan Turing Building, University of Manchester, Oxford Road, M13 9PL, UK\\
$^{2}$Leiden Observatory, Leiden University, P.O. Box 9513, NL-2300 RA Leiden, The Netherlands\\
$^{3}$Astrophysics Sub-Department, Department of Physics, University of Oxford, Denys Wilkinson Building, Keble Road, Oxford OX1 3RH, UK\\
$^{4}$SETI Institute, 339 Bernardo Avenue, Suite 200, Mountain View, CA 94043, USA\\
$^{5}$Berkeley SETI Research Center, University of California, Berkeley, CA 94720, USA\\
$^{6}$University of Malta, Institute of Space Sciences and Astronomy, Msida, MSD2080, Malta\\
}

\date{28th November 2024}

\pubyear{2024}

\begin{document}

\label{firstpage}
\pagerange{\pageref{firstpage}--\pageref{lastpage}}
\maketitle

\begin{abstract}
The Atacama Millimeter/Submillimeter Array (ALMA) remains unparalleled in sensitivity at radio frequencies above 35 GHz. In this paper, we explore ALMA's potential for narrowband technosignature detection, considering factors such as the interferometer's undistorted field of view, signal dilution due to significant drift rates at high frequencies and the possibility of spectral confusion. We present the first technosignature survey using archival ALMA data in Band 3, focusing on two spectral windows centred on 90.642 GHz and 93.151 GHz. Our survey places new limits at these frequencies on the prevalence of extraterrestrial transmitters for 28 galactic stars, selected from the \textit{Gaia} DR3 catalogue. We employ a stellar 'bycatch' method to sample these objects within the undistorted field of view of four ALMA calibrators. For the closest star in our sample, we find no evidence of transmitters with $EIRP_{min} > 7 \times 10^{17}$ W. To the best of our knowledge, this represents the first technosignature search conducted using ALMA data.


\end{abstract}




\section{Introduction}
The search for extraterrestrial intelligence (SETI) has yet to detect conclusive evidence of intelligent life beyond Earth. Despite this, there are continuing efforts to expand the volume of space surveyed. Increasingly sensitive radio telescopes (in particular through the use of distributed arrays) have provided the opportunity to detect unintentional leakage from civilisations within our vicinity, as well as searching for high-powered beacons detectable on galactic scales. The Breakthrough Listen project \citep{worden_breakthrough_2017} has also transformed the field with the installation of state-of-the-art commensal backends and dedicated observing time enabling SETI surveys to be conducted at 100-m Robert C. Byrd Green Bank Telescope, 64-m CSIRO Parkes telescope, MeerKAT and most recently the VLA via the COSMIC project (see \citet{enriquez_breakthrough_2017}, \citet{price_breakthrough_2018}, \citet{czech_breakthrough_2021} \& \citet{tremblay_cosmic_2023}). 

Electromagnetic waves at radio frequencies remain optimum for interstellar communication since they travel at the speed of light and propagate unimpeded through interstellar dust \citep{cocconi_searching_1959}. Radio technosignatures include both intentional beacons and unintentional communication leakage from an extraterrestrial civilisation; technosignatures are most easily discernible if they are  narrowband in frequency and therefore distinct against the natural broadband emission associated with our own galaxy and more distant cosmic radio sources. Other potential forms of technosignatures, such as negatively-dispersed broadband signals \citep{gajjar_searching_2022} or energy-efficient pulsating beacons \citep{cullers_sensitive_1986}, remain challenging profiles to search for due to computational limitations. SETI also considers the potential for technosignature detection at infrared \citep{townes} and optical wavelengths \citep{optical_shelley}, especially in the search for waste heat leakage through anomaly detection \citep{zuckerman}.

A technosignature, for successful detection, must overcome noise from both the sky background and the telescope receivers themselves \citep{kardashev_transmission_1964}. The sky background is a minimum at frequencies around a few GHz - at lower frequencies, the sky becomes very bright due to galactic emission and at higher frequencies, the cosmic microwave background and a receiver's quantum noise become dominant.  In addition, molecular absorption within the Earth's atmosphere becomes problematic above 10~GHz \citep{steffes_potential_1993}. Searches for extraterrestrial life have therefore been restricted to the ‘water hole’, defined as the frequency range bounded by the Hydrogen line at 1.42 GHz and Hydroxyl emission at 1.660~GHz \citep{oliver_project_1971}. For decades, SETI researchers have assumed that other technical civilisation would also consider this a special part of the electromagnetic spectrum since H and OH are the dissociation products of water, and water is recognised as essential to the development of life as we know it \citep{oliver_rationale_1979}. 

To date, SETI surveys have therefore concentrated on L-Band observations, with more recent efforts expanding the search up to 20~GHz (see \citet{choza_radio_2024}, \citet{price_breakthrough_2020}, \citet{sheikh_breakthrough_2020} and \citet{shirai_search_2004}). Besides a small number of surveys centred around the spectral line for positronium at 203~GHz (see \citet{steffes_deboer} \& \citet{mauersberger_seti_1996}), the millimetre and submillimetre band is largely unexplored.

Radio frequency interference (RFI) continues to hamper single-dish SETI observations, with false positives dominating data analysis, especially at cm-wavelengths. Long-baseline interferometry suppresses the response of RFI and also increases robustness to ETI signal detection \citep{garrett_seti_2018}. Early VLBI SETI surveys - \citet{rampadarth_vlbi} \& more recently \citet{wandia_interferometric_2023} - demonstrate interferometry's capacity for both targeted and commensal surveys.  

We also note that the impact of the ionised interstellar medium (ISM) and interplanetary medium (IPM) on technosignature searches is significantly reduced at millimetre and submillimetre wavelengths. At centimetre wavelengths, such as those around the water hole, temporal and spectral broadening, as well as other propagation effects caused by ionised gas, can distort and broaden extraterrestrial signals significantly \citep{Cordes_1993}. In principle, intrinsically narrowband signals artificially generated at millimetre and submillimetre wavelengths with milli-Hz bandwidths, should remain detectable as narrow drifting sub-Hz signals, even after propagating across the ISM of the Galaxy.  This allows much larger signal processing gains to be achieved, as one can productively channelize to finer bandwidths \citep{brzycki_detecting_2023}.

The emergence of large public data archives also permits SETI to be conducted in a commensal mode, re-utilising data observed for other astronomical projects. For any given field of view, a sample of stars \citep{wlodarczyk-sroka_extending_2020} and exotic objects \citep{garrett_constraints_2023} will always be present, and this permits much tighter constraints to be placed on the prevalence of extraterrestrial transmitters, such as by \citet{wlodarczyk-sroka_extending_2020}.

The Atacama Large Millimeter/Submillimeter Array (ALMA) provides a high resolution window on the universe from 35 to 950~GHz. Located in the Chajnantor Plateau, ALMA offers extensive coverage across the radio spectrum above 35~GHz due to minimal water vapour being present at this altitude. The main interferometer is a homogeneous array consisting of 54 12-m dishes, each antenna fitted with 10 receiver bands (at present, 8 receivers are operational, with Band 1 in its first cycle of observations and Band 2 still under development).

ALMA's current science goals focus on understanding the origins of the universe and continues to drive research in the fields of protoplanetary disks, galaxy evolution/kinematics and the polarisation of dust. So far as we are aware, no SETI observations have been conducted by ALMA. Shifting SETI studies towards millimetre and submillimetre wavelengths using ALMA offers several obvious advantages: (i) the RFI environment is vastly superior to that at cm-wavelengths, (ii) the sensitivity of the ALMA interferometer at these wavelengths is unparalleled, and (iii) it opens up a new and potentially very important window for SETI observations, significantly broadening the surveyed parameter space.   

In this paper, we describe the first SETI results obtained from data generated by the ALMA telescope. In Section 2, we outline the challenges associated with conducting SETI at these frequencies. Section 3 details the ALMA data we have retrieved from the archive and the stellar sample associated with these observations. Section 4 presents the results of a search for narrowband signals in these data, focusing on 28 stars located within the field of view of 4 calibrator scans. In Section 5, we place limits on the prevalence of extraterrestrial transmitters and compare this work against other SETI surveys. In Section 6, we present our conclusions.

\section{Motivation for High Frequency SETI} 
There are at least three main issues that need to be considered before embarking on a SETI search with a high frequency and high resolution interferometer such as ALMA. These include (i) field of view of the interferometer array, (ii) the high drift rate of narrowband signals at millimetre and sub-millimetre wavelengths and (iii) the potential for spectral confusion. We consider these three issues in turn.

\subsection{Antenna Primary Beam Width and Interferometer Field of View }
The field of view of radio telescopes at millimetre and submillimetre wavelengths are usually significantly smaller than telescopes operating at longer wavelengths. For SETI, the field of view is an increasingly important part of discovery space and it is especially crucial for studies that make use of archival data - a large field of view maximises the number of stars that are potential targets in a SETI analysis. 

The field of view of ALMA is dependent on various factors, including the Full Width at Half Maximum (FWHM) of the individual antenna primary beam $\theta_{PB}$: 
\begin{equation}
\label{eq:fov}
\theta_{PB} = \frac{1.22\lambda}{D}
\end{equation}
where $D$ is the diameter of a single dish and $\lambda$ is the observing wavelength. 

The FWHM of the primary beam sets the ultimate field of view for a telescope. However, for interferometers like ALMA, other factors such as bandwidth and time-average smearing must also be considered. For a given array configuration and observing frequency $\nu_0$,  the severity of these effects depends on the bandwidth ($\delta\nu$) and integration time ($\tau_{av}$) of the correlated data - poor frequency and time resolution will suppress the interferometer's sensitivity to sources located towards the edge of the field of view \citep{wrobel_vlbi_1995}. The better the time and frequency resolution, the larger the undistorted field of view. The severity of bandwidth and time-average smearing will limit the number of stars located within an interferometer's instantaneous field of view. 

The undistorted field of view due to bandwidth smearing, $\theta_{bw}$, is based on the spectral resolution $\delta\nu$ and central observing frequency $\nu_0$. The undistorted field of view is defined as the limit within which smearing results in less than a 10\% reduction in the point source response, such that,
\begin{equation}
\label{eq:bwsmear}
\theta_{bw} \approx 0.8 \frac{\nu_0}{\delta\nu} \theta_{syn}
\end{equation}

where $\theta_{syn}$ is the width of the synthesised beam \citep{wrobel_vlbi_1995}. 

The synthesised beam, $\theta_{syn}$, depends on the maximum baseline length $B_{max}$ and can be estimated by equation \ref{eq:syn},
\begin{equation}
\label{eq:syn}
\theta_{syn} = \frac{c}{B_{max}\nu}
\end{equation}

For all practical purposes, the large fractional bandwidth of ALMA $\frac{\nu_0}{\delta\nu} > 3 \times 10^6$ means that frequency smearing is not a major limitation on the interferometer's field of view at any frequency band or in any array configuration. 

Time-average smearing is a more serious effect. The limit of distortion due to time-average smearing, $\theta_{t}$, resulting in a 10\% reduction in the point source response is,
\begin{equation}
\label{eq:tsmear}
\theta_{t} \approx 9000 \frac{\theta_{syn}}{\tau_{av}}
\end{equation}
where $\tau_{av}$ is the correlator averaging time \citep{wrobel_vlbi_1995}. 

The maximum field of view due to time-average smearing is presented in Table \ref{table:1}, assuming a typical correlator integration time of 6~s and the longest ALMA baseline length of 16~km. The field of view defined by the primary beam $\theta_{PB}$ and time smearing $\theta_{t}$ are very similar; time-average smearing does not notably diminish ALMA's sensitivity across the antenna's primary beam. In a SETI analysis using archival data, this implies that ALMA can detect signals from other stars across the FWHM of the antenna's primary beam.

\begin{table}
\setlength{\arrayrulewidth}{0.2mm}
\setlength{\tabcolsep}{15pt}
\renewcommand{\arraystretch}{1.5}
\centering
\caption{Estimates of the antenna primary beam size and undistorted interferometer field of view are presented for ALMA Band 3, 6 and 10.}
\begin{tabular}{c|c|c|c} 
 $\nu$ (GHz) & $\theta_{PB}$ (arcsec) & $\theta_t$ (arcsec)\\ [0.5ex] 
 \hline
 100 & 62.9  & 58.5 \\ [1ex] 
 250 & 25.2 & 22.5 \\ [1ex] 
 800 & 7.8 &  7.5 \\ [1ex] 
\end{tabular}
\label{table:1}
\end{table}

\subsection{Drift Rate at High Frequencies}\label{subsec_driftrate}
An artificial narrowband signal will change its central frequency due to the relative acceleration between the transmitter and receiver \citep{sheikh_choosing_2019},

\begin{equation}
    \dot{\nu} = \frac{d v_r}{dt} \frac{\nu_{\text{rest}}}{c}
\end{equation}

where $\dot{\nu}$ is the drift rate, $\nu_{\text{rest}}$ is the rest frequency of the signal, and $\frac{d v_r}{dt}$ is the total relative radial acceleration between the transmitter and receiver.



The time taken for a drifting signal to pass through a channel bandwidth of $\delta\nu$ is
\begin{equation}
\label{eq:time}
t = \frac{\delta\nu}{\dot{\nu}}
\end{equation}
where $\delta\nu$ is the channel bandwidth, in units of Hz. 

The maximum drift rate $\dot{\nu}$ of a narrowband SETI signal will depend on the frequency observed, $\nu_0$ (in GHz), such that, 

\begin{equation}
\label{eq:drift2}
\dot{\nu}_{max} = \dot{\nu}_{\rm 1GHz} \times \nu_0/\rm{1~GHz}
\end{equation}
where $\dot{\nu}_{max}$ is the maximum drift rate (Hz/s) at observing frequency $\nu_0$ and $\dot{\nu}_{\rm 1GHz}$ the maximum drift rate in units of Hz/s at 1~GHz. We assume maximum drift rates of $\pm 4$~Hz/s following the current Breakthrough Listen campaigns made at L-band \citep{price_breakthrough_2020}. The maximum drift rate of a signal therefore scales with frequency multiplicatively, resulting in drift rates 1000 times larger in ALMA Band 10 than for signals typically observed at 1~GHz (within the water hole). The rapidly scaling drift rate at high radio frequencies is a potential problem for SETI observations. 

From equation \ref{eq:drift2}, measurements made using ALMA's Band 3 would have a maximum drift rate of $\sim \pm 400$~Hz/sec. For the minimum channel bandwidth of 30~kHz in our archive data set, equation \ref{eq:time} predicts that a narrowband signal will cross frequency channels on time scales of $\sim 75$ seconds. Therefore, rapidly-drifting signal dilution can be mitigated by integrating over timescales less than 75 seconds for Band 3; this timescale will decrease with increasing frequency.

\subsection{Spectral Confusion} 
Spectral confusion is a significant consideration when using ALMA for astronomical observations, including SETI. ALMA operates across a wide range of frequencies, where the presence of numerous spectral lines from natural astrophysical sources can complicate data interpretation. These spectral lines arise from various molecular and atomic transitions in interstellar and circumstellar environments, contributing to a dense spectral landscape above 100 GHz (see \citet{splatalogue} \& \citet{mcguire_2021_2022}). Efforts to mitigate spectral confusion in archival data could include a careful selection of target sources that are not expected to show complex spectral line emission, e.g. extragalactic calibration sources. However, it's important to note that even at millimetre and submillimetre wavelengths, SETI narrowband radio technosignatures are expected to be much narrower in frequency width than any naturally occurring spectral line emission.

\section{Observations of the Stellar Sample \& Initial Data Analysis}

Given the previous discussion, we searched the ALMA archive for observations made with the very highest spectral resolution. We also focused on data that included continuum calibrator scans associated with targets lying relatively close to the Galactic Plane. This maximises the stellar bycatch within the field of view. Observations were restricted to galactic latitudes of $|b| < 5^\circ$ and $\delta \nu < 35$~kHz. Mosaic observations were avoided as their targets were often associated with spatially distributed and complex molecular cloud regions. We decided to focus our attention on Band 3 observations (84 - 116~GHz), as these had the largest field of view ($53" - 74"$).

Eventually, we converged on ALMA project 2017.1.01794.S, a Band 3 observation of 12 star-forming clumps. This observation employed very high spectral resolution in order to observe molecular tracers N2H \& HNC in two spectral windows (90.608 - 90.667~GHz) and (93.117 - 93.175~GHz). The data were correlated with 3840 frequency points, each channel being 30.52 kHz wide. This project used a configuration of 45 antennas, with the 5th percentile baseline length of 24~m and a maximum baseline of 314~m. 

\begin{figure*}
\centering
\begin{subfigure}{0.48\linewidth}
\includegraphics[width=\linewidth]{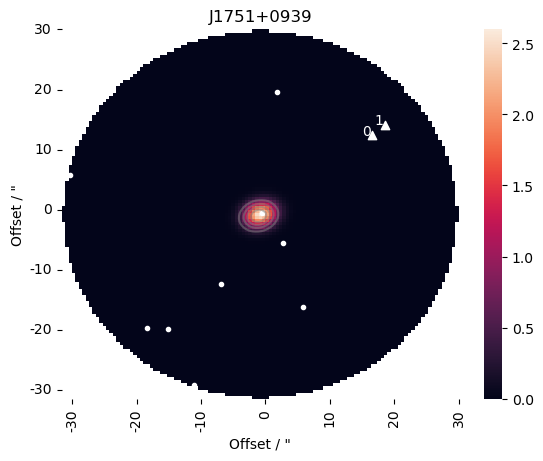}
\label{fig:J1751}
\end{subfigure}
\begin{subfigure}{0.48\linewidth}
\includegraphics[width=\linewidth]{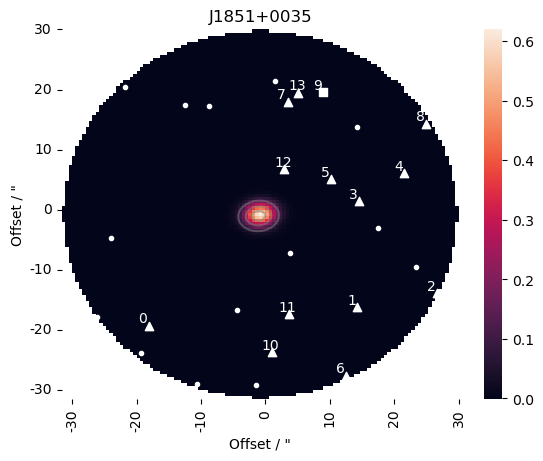}
\label{fig:J1851}
\end{subfigure}
\begin{subfigure}{0.48\linewidth}
\includegraphics[width=\linewidth]{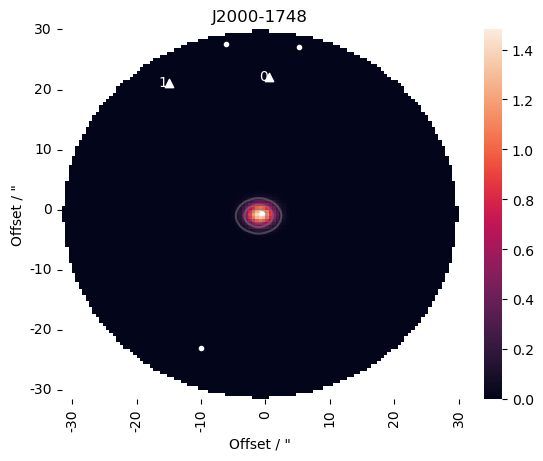}
\label{fig:J2000}
\end{subfigure}
\begin{subfigure}{0.48\linewidth}
\includegraphics[width=\linewidth]{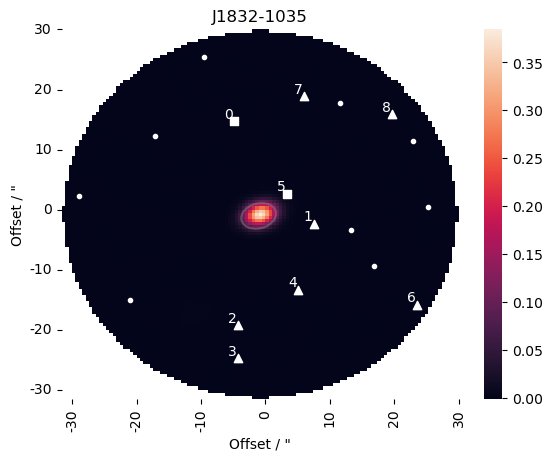}
\label{fig:J1832}
\end{subfigure}
\caption{The stellar samples located within each 59 arcsecond field of view superimposed on clean images of the four calibrators associated with spectral window $90.608-90.666$~GHz. The sample of stars surveyed include stars with distances calculated by the inverse parallax (square markers) or geometric distance estimations from Bailer-Jones (triangle markers). Stars with circular markers have a poor fractional parallax error and were not included in the sample. Stars are numbered based on the Gaia DR3 Source ID, corresponding to the star properties given in Table \ref{tab:star_properties}.}
\label{fig:cal_fov}
\end{figure*}

We performed a search of the \textit{Gaia} DR3 catalogue (Gaia Collaboration et al. \citeyear{gaia_collaboration_gaia_2016}, \citeyear{gaia_collaboration_gaia_2023}) using \textsc{astroquery} to identify stars located within the 59" field of view.  Figure \ref{fig:cal_fov} presents the positions of these stars relative to the central calibrator and the reliability of their distance estimates.  The accuracy of the distance estimate for each star is dependent on the fractional parallax uncertainty ($f = \frac{\sigma_\omega}{\omega}$). For stars with 0 < f < 0.2, the distance to the star is the inverse of the parallax. For stars with 0.2 < f < 1, \citet{bailer-jones_estimating_2021} provides geometric distance estimations for stars in \textit{Gaia} DR3. Stars with f < 0 or f > 1 were excluded from our sample. The basic properties of the stars in our sample, including their estimated distances, are presented in Table \ref{tab:star_properties}. 

\begingroup
\setlength{\arrayrulewidth}{0.2mm}
\renewcommand{\arraystretch}{1.5}
\begin{table*}
\caption{Properties of stars surveyed and their associated $EIRP_{min}$. Estimations of the $EIRP_{min}$ were based from the median r.m.s noise within each field of 27~mJy (J1751+0939), 38~mJy (J1851+0035), 25~mJy (J200-1748) \& 37~mJy (J1832-1035). The stars have been numbered based on Gaia DR3 Source ID.}
\label{tab:star_properties}
\begin{tabular}{|lllllllll} 
\hline
Field & No. & Gaia DR3 Source ID & Right Ascension & Declination & Magnitude & Distance & $EIRP_{min}$\\
& & & (hms) & (dms) & & (kpc) & (W)\\
\hline
J1751+0939 & 0 & 4488787487759680128 & 17:51:31.802 & +09:39:13.308 & 19.016 & 3.037 & 4.55 $\times 10^{18}$\\
 & 1 & 4488787492055460992 & 17:51:31.673 & +09:39:14.968 & 19.576 & 2.126 &  2.23 $\times 10^{18}$\\ 
J1851+0035 & 0 & 4266512759708638208 & 18:51:47.961 & +00:35:14.637 & 19.415 & 4.843 & 1.63 $\times 10^{19}$\\
& 1 & 4266512759708640896 & 18:51:45.871 & +00:35:17.610 & 19.674 & 3.064 & 6.52 $\times 10^{18}$\\
& 2 & 4266512759708643968 & 18:51:45.073 & +00:35:19.904 & 18.777 & 3.787 & 9.96 $\times 10^{18}$\\
& 3 & 4266512759708662912 & 18:51:45.851 & +00:35:34.722 & 18.444 & 5.165 & 1.85 $\times 10^{19}$\\
& 4 & 4266512759708669440 & 18:51:45.398 & +00:35:39.214 & 18.146 & 3.491 & 8.46 $\times 10^{18}$\\
& 5 & 4266512759715601408 & 18:51:46.136 & +00:35:38.292 & 20.677 & 3.757 & 9.80 $\times 10^{19}$\\
& 6 & 4266512764004977280 & 18:51:45.986 & +00:35:06.588 & 20.498 & 4.298 & 1.28 $\times 10^{19}$\\
& 7 & 4266512798366410624 & 18:51:46.569 & +00:35:50.826 & 19.895 & 3.438 & 8.21 $\times 10^{18}$\\
& 8 & 4266512832730076672 & 18:51:45.176 & +00:35:47.238 & 16.421 & 1.401 & 1.36 $\times 10^{18}$\\
& 9 & 4266512862787902208 & 18:51:46.213 & +00:35:52.284 & 16.491 & 
1.102 & 8.44 $\times 10^{17}$\\
& 10 & 4266512764010583936 & 18:51:46.726 & +00:35:10.391 & 17.505 & 1.416 & 1.39 $\times 10^{18}$\\
& 11 & 4266512764010585984 & 18:51:46.556 & +00:35:16.526 & 17.675 & 3.025 & 6.35 $\times 10^{18}$\\
& 12 & 4266512764010592512 & 18:51:46.600 & +00:35:39.936 & 16.480 & 3.792 & 9.98 $\times 10^{18}$\\
& 13 & 4266512794075341440 & 18:51:46.460 & +00:35:52.262 & 19.398 & 3.822 & 1.01 $\times 10^{19}$\\
J2000-1748 & 0 & 6867841551725448832 & 20:00:57.123 & -17:48:36.224 & 16.665 & 2.939 & 3.96 $\times 10^{18}$\\
& 1 & 6867841551726377984 & 20:00:58.176 & -17:48:37.290 & 19.255 & 2.218 & 2.26 $\times 10^{18}$\\
J1832-1035 & 0 & 4154920816345457536 & 18:32:21.224 & -10:34:55.946 & 13.155 & 2.694 & 4.91 $\times 10^{18}$\\
& 1 & 4154920816349239168 & 18:32:20.412 & -10:35:12.462 & 18.850 & 4.702 & 1.50 $\times 10^{19}$\\
& 2 & 4154920820643410176 & 18:32:21.185 & -10:35:28.802 & 20.061 & 6.025 & 2.46 $\times 10^{19}$\\
& 3 & 4154920820659120000 & 18:32:21.186 & -10:35:34.198 & 18.235 & 2.032 & 2.79 $\times 10^{18}$\\
& 4 & 4154920820659124352 & 18:32:20.571 & -10:35:23.106 & 17.992 & 1.839 & 2.29 $\times 10^{18}$\\
& 5 & 4154920820659128192 & 18:32:20.683 & -10:35:07.615 & 16.606 & 1.010 & 6.91 $\times 10^{17}$\\
& 6 & 4154920820659128320 & 18:32:19.356 & -10:35:25.518 & 17.501 & 1.873 & 2.38 $\times 10^{18}$\\
& 7 & 4154920820659131520 & 18:32:20.510 & -10:34:51.911 & 17.166 & 1.579 & 1.69 $\times 10^{18}$\\
& 8 & 4154921576557696768 & 18:32:19.611 & -10:34:54.803 & 19.821 & 1.965 & 2.61 $\times 10^{18}$\\
\hline
\end{tabular}
\end{table*}
\endgroup

Data were taken during five observing runs, from 2nd July 2017 and 12 July 2017, using the 12m array. For four observational runs, J1751+0939 was used as the flux and bandpass calibrator and J1851+0035 was used as the phase calibrator. For the fifth observational run, J2000-1748 was used as the flux and bandpass calibrator and J1832-1035 was used as a phase calibrator. Initial sampling constraints along the Galactic Plane apply to the phase calibrators located in close proximity to the target fields; the flux calibrators do not necessarily lie within latitudes of $|b| < 5^\circ$ but have been included nevertheless.

\begin{table*}
 \caption{Properties of four targeted calibrators. The flux density of each calibrator is taken from the most recent observation taken at Band 3 (91.5~GHz) in the ALMA Calibrator Catalogue.}
 \label{tab:calibrators}
 \begin{tabular}{ccccccc}
  \hline
  Calibrator & Type & RA & Dec & Redshift & Flux Density & Cubes Produced\\
   & & (hms) & (dms) & & (Jy) &\\
  \hline
  J1751+0939 & BL Lac & 17:51:32.819 & +09:39:00.728 & 0.322 & 2.90 & 5 $\times$ 60s\\[2pt] 
  J1851+0035 & Radio & 18:51:46.723 & +00:35:32.365 & - & 0.43 & 7 $\times$ 30s \\[2pt]
  J2000-1748 & QSO & 20:00:57.090 & -17:48:57.673 & 0.652 & 2.60 & 5 $\times$ 60s\\[2pt]
  J1832-1035 & QSO & 18:32:20.836 & -10:35:11.197 & - & 0.54 & 7 $\times$ 30s\\[2pt]
  \hline
 \end{tabular} 
\end{table*}

ALMA calibrators make ideal candidates for commensal SETI research as they are usually bright continuum sources with simple structure and accurate positions (see also section 2.3). The properties of these calibrators are presented in Table \ref{tab:calibrators}. 

The associated visibility data were downloaded from the ALMA Archive and calibrated using the ALMA pipeline in CASA 5.6, as recommended for Cycle 7 data. The calibrators followed the same standard interferometric calibration procedure as the science target fields, correcting for fluctuations in system temperature, antenna position and water vapour radiometer measurements. The pipeline derives a bandpass correction from the flux calibrators and phase offsets in the spectral windows are also derived. These calibration tables are also applied to the phase calibrators. The sources were also self-calibrated with a point source model using flux densities interpolated from ALMA's Calibrator Catalogue.  Two rounds of self-calibration are performed - phase only and then amplitude only.  

We attempted further self-calibration of the calibrators but could not achieve any significant improvement beyond the pipeline calibration. Continuum images of the calibrators were made using CASA 6.5 and visually inspected. The clean images are shown in Fig.~\ref{fig:cal_fov}. The r.m.s noise in the continuum maps, ranging from 280-340 $\mu$Jy/beam, is consistent with theoretical estimates.

\section{SETI Data Analysis \& Results}

We subtracted clean component models of the continuum sources from the uv-data using CASA task \textit{uvsub}. To avoid the dilution of a drifting narrowband signal across multiple frequency channels, the calibrator subtracted uv-data were split into multiple data sets with total scan lengths limited to < 75 seconds. The flux calibrators, J1751+0939 and J2000-1748, were observed for 5 minutes each and were therefore separated into 5 x 1 minute data sets. The phase calibrators, J1851+0035 and J1832-1035, were each observed for 30 seconds on 7 separate occasions, separated by $\sim 10$ minutes.

\begin{figure}
\centering
\includegraphics[width=0.9\linewidth]{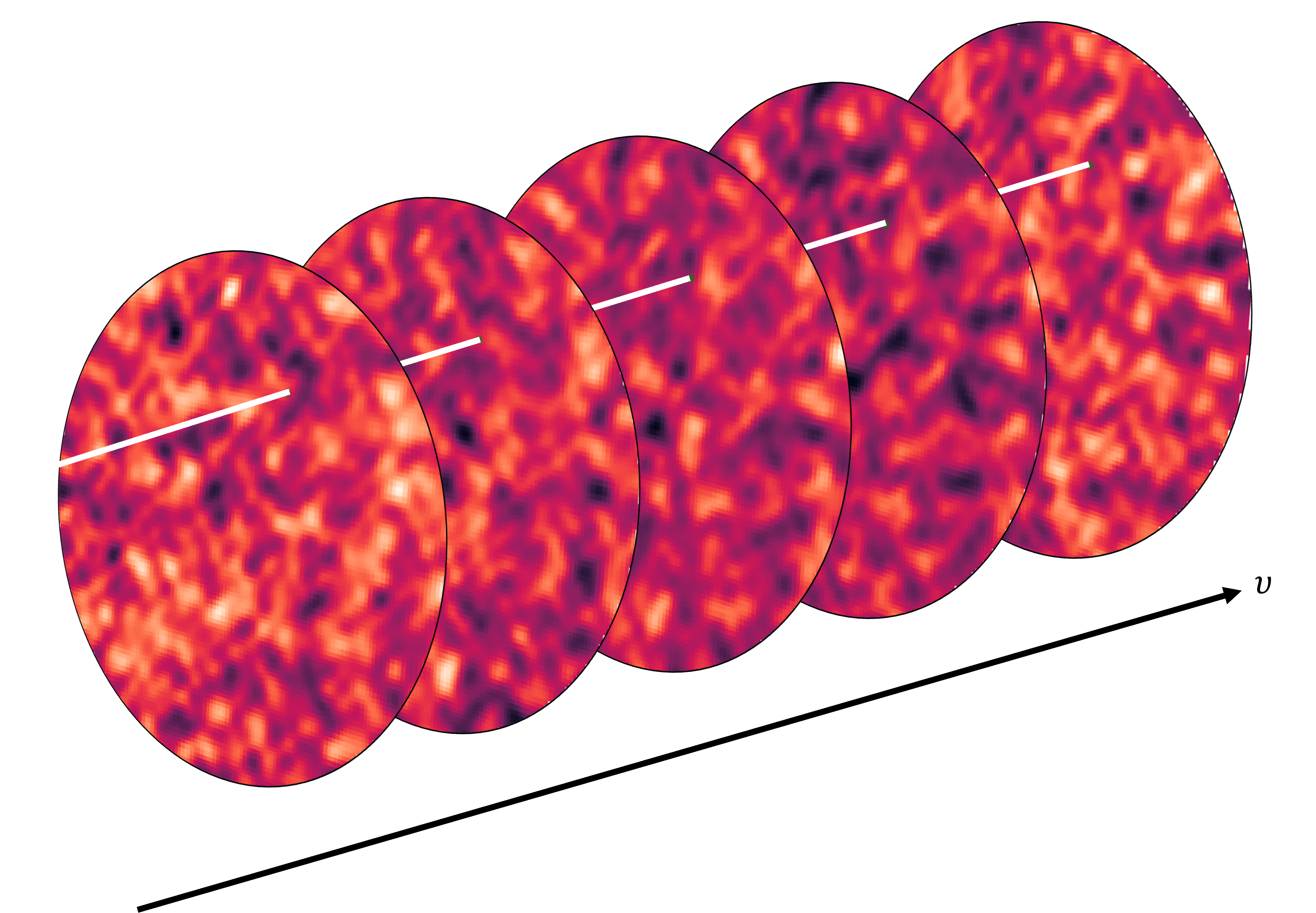}
\caption{Frequency slices of the calibrator-subtracted data cube for J1751+0939, highlighting the pixel containing Gaia DR3 4488787487759680128. The signal-to-noise ratio is calculated for each channel map for a given pixel.}
\label{fig:slice}
\end{figure}

For each data set, channel maps were generated and stored as image cubes - no cleaning was performed on the data. The r.m.s noise level for each channel map was estimated using CASA's \textit{imstat}. For example, the r.m.s noise levels we measured within a single channel map is $\sim 34$~mJy/beam. Fig. \ref{fig:slice} presents a visualisation of a data cube (in this case for J1751+0939) showing some of the channel maps and the pixel associated with Gaia DR3 4488787487759680128.   

\begin{figure}
\centering
\includegraphics[width=\linewidth]{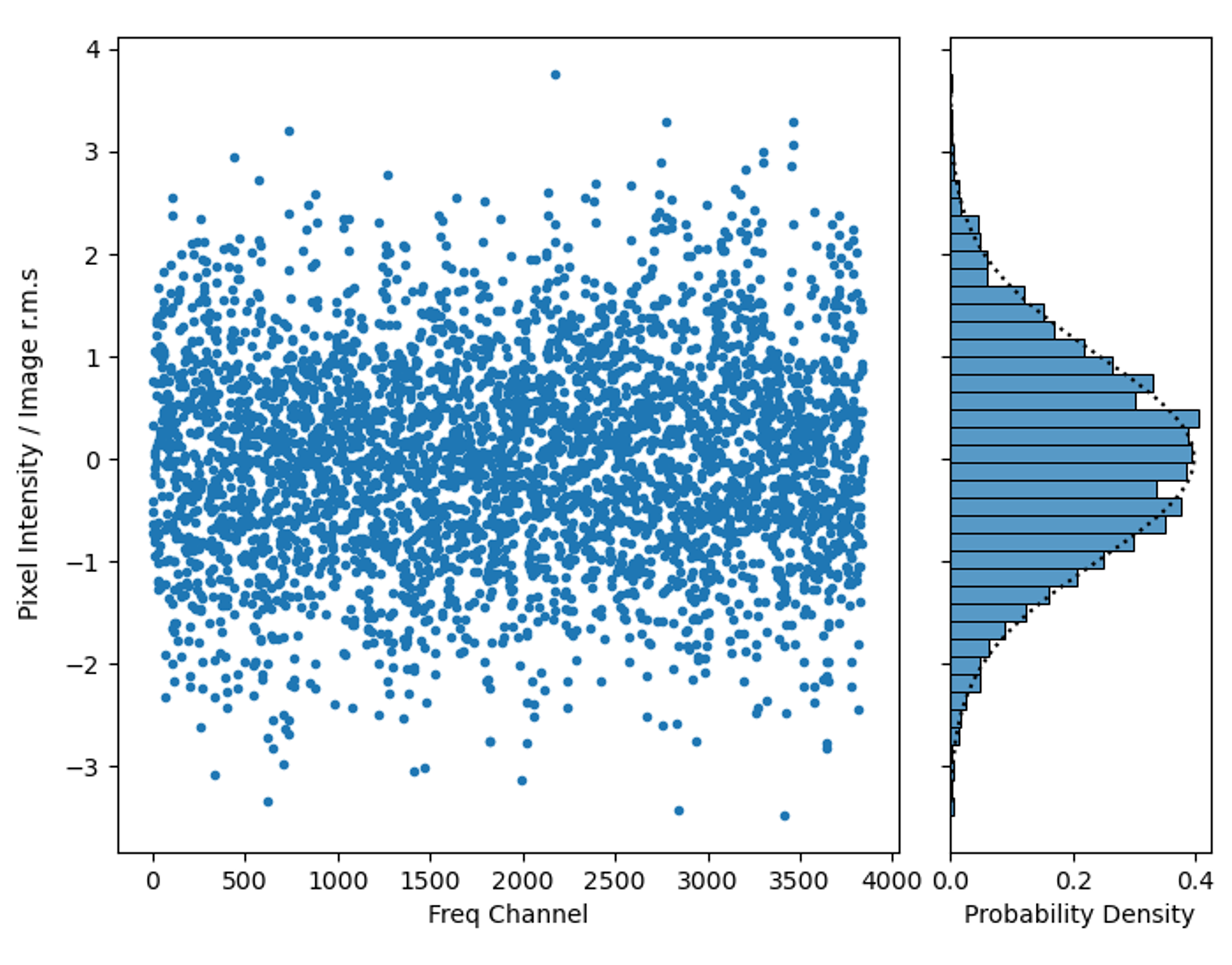}
\caption{The flux:r.m.s noise ratio for one of the stars in our sample, Gaia DR3 4488787487759680128 plotted as a function of channel frequency and the corresponding signal-to-noise histogram. Random emission from thermal noise will result in a Gaussian distribution of intensities, whereas a potential ETI signal would have a $SNR > 5$.}
\label{fig:scatter_hist}
\end{figure}

For each pixel corresponding to a star in our sample and for each channel map, the ratio of the pixel intensity to image r.m.s was estimated, as shown in Figure \ref{fig:scatter_hist}. We assumed a Gaussian distribution for a pixel across all frequency channels and used $SNR > 5$ as our detection threshold.   

Across the 120 data cubes, no signals were detected with a $SNR > 5$. We also search for signals in adjacent pixels, covering a diameter of 1.1" centred on each star in our sample. No signals with a $SNR > 5$ were observed in these surrounding regions. We calculated the associated $EIRP_{min}$ for each star in our sample, 
\begin{equation}
\label{eq:eirp}
EIRP_{min} = 4 \pi d^2 S_{min} \delta\nu
\end{equation}
where d is the distance to the source, $S_{min}$ is minimum detectable flux density and $\delta\nu$ is the spectral resolution. We estimated the minimum detectable flux density, $S_{min}$, as the SNR threshold multiplied by the r.m.s measured within the field of view of each calibrator. These limits are presented for our sample of stars in Table \ref{tab:star_properties}. 
For assumed isotropic emission, the smallest minimum detectable power ($EIRP_{min}$) our measurements are sensitive to is $6.91 \times 10^{17}$~W for the closest target star (Gaia DR3 4154920820659128192). The sensitivity of this survey assumes the technosignature is continuous in nature; a pulsed signal would still be detectable with our method, but would be diluted by a factor depending on the duty cycle of the signal.

In summary, we detect no signals with an $EIRP_{min} > 6.91 \times 10^{17}$~W. 

\section{Discussion}
Clearly, ALMA opens up a new and important area of parameter space for SETI studies. We have demonstrated that archival ALMA data can be used to place interesting constraints on the prevalence of extraterrestrial transmitters at millimetre wavelengths. 

\begin{figure*}
\centering
\includegraphics[width=\linewidth]{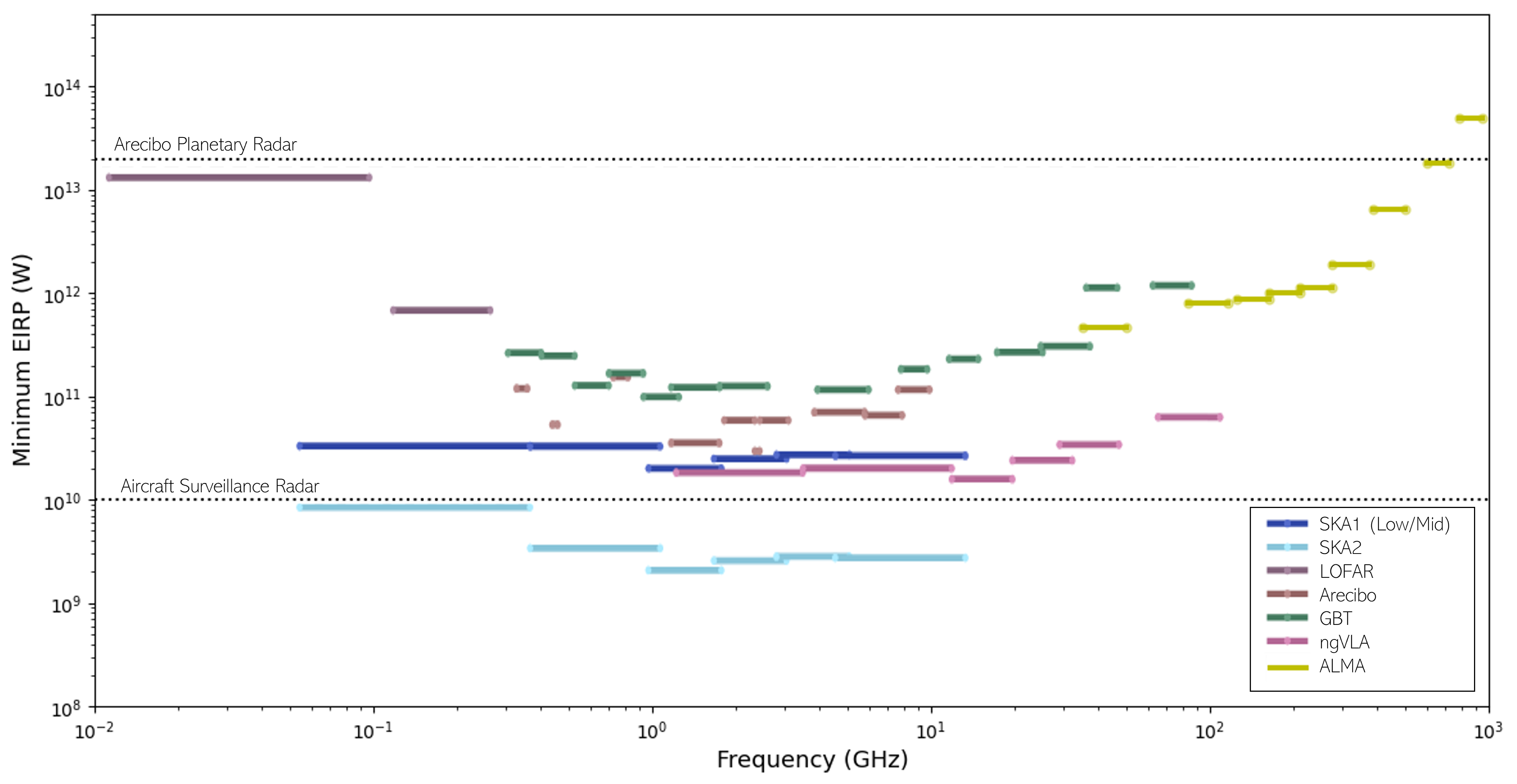}
\caption{Comparison of the potential sensitivity of telescopes to perform SETI searches for narrowband transmitters against frequency coverage (following  \citet{siemion_searching_2014} \& \citet{croft_science_2018}). The telescope's minimum detectable $EIRP_{min}$ is based on an integration time of 10~minutes, a significance threshold of 15, channel width of 0.5~Hz and distances up to 15~pc. ALMA operates at a higher frequency range than previously explored and is sensitive to radio powers similar to the Arecibo planetary radar system.}
\label{fig:eirp}
\end{figure*}

It is interesting to compare the sensitivity of ALMA with other telescopes currently or anticipated to be involved in SETI surveys. Following \citet{siemion_searching_2014} \& \citet{croft_science_2018}, Fig. \ref{fig:eirp} presents each telescope's $EIRP_{min}$ in a given frequency band. We assume an integration time of 10~minutes, a channel width of 0.5~Hz, a source distance of 15~pc and a SNR > 15. Clearly, ALMA significantly expands the potential frequency coverage for detecting narrowband technosignatures by over an order of magnitude and with good sensitivity.

While ALMA's raw sensitivity aligns well with other SETI facilities, achieving Hz spectral resolution would necessitate a specialised backend, similar to that of the COSMIC VLA project \citep{tremblay_cosmic_2023} or the BLUSE system on MeerKAT \citep{czech_breakthrough_2021}. A beamforming approach would be required in order to keep the output data rates manageable. 

It is also interesting to consider the merit of this archival investigation using ALMA, in comparison with other SETI surveys. We use the continuous wave transmitter figure of merit (CWFTM) as defined by \citet{enriquez_breakthrough_2017} and adopted thereafter by many other authors,
\begin{equation}
\label{eq:cwtfm}
CWTFM = \zeta_{A0} \frac{EIRP_{min}}{N_* \nu_{frac}}
\end{equation}
where $N_*$ is the number of stars surveyed and $\nu_{frac}$ is the ratio of the total bandwidth to the central frequency, $\frac{\Delta\nu}{\nu_0}$. Note that $\zeta_{A0}$ is the normalisation constant such that CWFTM = 1 relates to a survey of 1000 stars with $\nu_{frac}$ = 0.5 to detect a transmitter of equivalent power to Arecibo ($10^{13}$~W). 

\begin{figure}
\includegraphics[width=\linewidth]{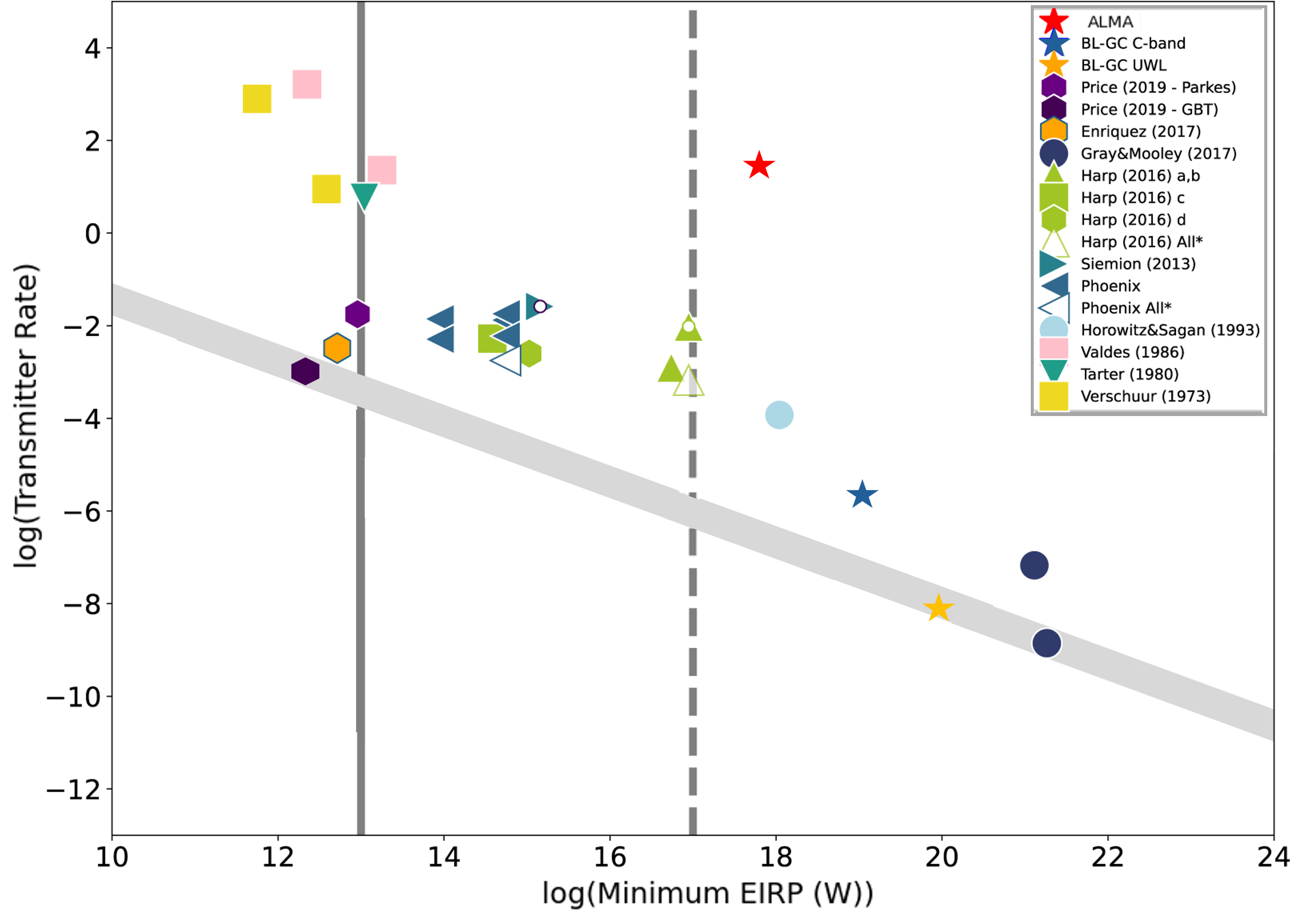}
\caption{The $EIRP_{min}$ and transmitter rate for a sample of SETI surveys above 1~GHz, including our own work using ALMA (amended from figure by \citet{tremblay_cosmic_2023}). The EIRP can be compared against potential limits for an Arecibo planetary radar ($10^{13}$W) and a Kardashev Type I civilisation ($10^{17}$W). The suggested 'Terra Incognito' limit - denoted in the slanted gray line - challenges the limits of telescopes and the breadth of surveys conducted to date.}
\label{fig:eirp_tr}
\end{figure}

Following \citet{enriquez_breakthrough_2017} and subsequent SETI papers, Figure \ref{fig:eirp_tr} plots the $EIRP_{min}$ against transmitter rate, $(N_*\nu_{rel})^{-1}$, for the archival investigation using 2017.1.01794.S against other SETI surveys.

It is important to note that the figure of merit of this paper's archival investigation is not fully indicative of ALMA's potential use as a SETI instrument, if it were equipped with a SETI backend system similar in performance to the COSMIC VLA or BLUSE MeerKAT systems. However, we also note that there are still some challenges for ALMA to overcome. Probably the biggest limitation is the large signal drift rates ($\pm 400$~Hz/s) that need to be corrected for in order to extend the coherence time of narrowband signals beyond very short integration times (see section 2.2). In particular, current algorithms can only deal with Doppler drift rates that are two orders of magnitude smaller than the values expected for ALMA \citep{sheikh_choosing_2019}. 

One possibility is to correct the data to a fiducial reference frame at a few hundred Hz/s and conduct searches across a small bandwidth within this frame. This follows the approach of the Mega-Channel Extraterrestrial Assay (META) \& Billion-Channel Extraterrestrial Assay (BETA) - see \citet{horowitz_meta} \& \citet{leigh_beta}. By correcting to different inertial frames across a bandwidth encapsulating the calculated doppler shift, it would be possible to mitigate smearing losses. While this approach does not reduce the overall computational burden, it does ensure that high drift rates can be effectively searched while preserving sensitivity. This strategy would enable high spectral resolution searches without a significant loss in detection performance, although it requires substantial computational resources to implement.

Another issue is the limited number of stars ($N_*$) with known distances located within the telescope's field of view. ALMA’s field of view is significantly smaller than that of the VLA (42' at 1~GHz) and MeerKAT (approx 1 square degree). We estimate that at 5.5 degrees above or below the Galactic Plane, the number of stars in the Gaia DR3 catalogue with good astrometric accuracy (0 < f < 1) is approximately 13 per square arcminute. We therefore expect approximately 10 stars within the field of view for Band 3 observations, but this is much reduced for Band 9 \& 10. Indeed, it is probable that no stars with Gaia distances fall within the FoV at these high frequencies. 

ALMA's use for SETI may therefore not challenge the 'terra incognito' of CWFTM plots \citep{garrett_constraints_2023} but ALMA's strength lies in the expansion to higher radio frequencies that have hitherto not been probed.

\section{Conclusion}

We have conducted the first SETI survey with ALMA using archival data from project 2017.1.01794.S. We identified 28 stars within the field of view of four calibrator fields using \textit{Gaia} DR3. We searched for signals with a SNR > 5 within the pixels containing these stars, and adjacent pixels up to 1.1" away.  No extraterrestrial transmitter signals were detected with an $EIRP_{min} > 6.91 \times 10^{17}$~W. Further progress can still be achieved through continuing a deeper analysis of archival interferometer data. The analysis of ALMA's archival data correlated with 35~kHz spectral resolution at significantly higher frequencies, such as in Band 6 and Band 7, may be promising.

We have presented the performance of ALMA as a SETI instrument, in comparison with other SETI surveys and telescopes, and suggest several steps for future ALMA-SETI exploration. There are several advantages to moving the search to higher frequencies in terms of propagation effects that distort and broaden narrowband signals at lower frequencies e.g. 1-2~GHz. Further progress with ALMA can be made via the the introduction of a specialised SETI backend to generate finer spectral resolution data and employing beamforming approaches to target galactic stars of known distance lying within the antenna field of view. Additional computational resources would also facilitate the implementation of more sophisticated Doppler correction techniques, such as correcting to a fiducial reference frame at a few hundred Hz/s and conducting searches within a small range around this frame. These advancements would enable deeper and more sensitive SETI searches using ALMA and other millimetre telescopes.

Finally, SETI surveys at millimetre and submillimetre wavelengths offer exciting opportunities to expand the parameter space explored for extraterrestrial intelligence. By demonstrating ALMA’s capability for SETI research, we encourage other high frequency technosignature surveys. We believe that ALMA presents a promising new avenue for SETI, unlocking an entirely new frequency window for technosignature detection.

\section*{Acknowledgements}
We would like to thank contributions from George Bendo, the UK ALMA Arc Node and ESO ALMA Helpdesk for assistance with ALMA data calibration and the use of computer resources.

\section*{Data Availability}
This paper makes use of the following ALMA data: ADS/JAO.ALMA 2017.1.01704.S ALMA is a partnership of ESO (representing its member states), NSF (USA) and NINS (Japan), together with NRC (Canada), MOST and ASIAA (Taiwan), and KASI (Republic of Korea), in cooperation with the Republic of Chile. The Joint ALMA Observatory is operated by ESO, AUI/NRAO and NAOJ. This work presents results from the European Space Agency (ESA) space mission \textit{Gaia}. \textit{Gaia} data are being processed by the \textit{Gaia} Data Processing and Analysis Consortium (DPAC). Funding for the DPAC is provided by national institutions, in particular the institutions participating in the \textit{Gaia} MultiLateral Agreement (MLA). The \textit{Gaia} mission website is https://www.cosmos.esa.int/gaia. The \textit{Gaia} archive website is https://archives.esac.esa.int/gaia. This research has made use of the VizieR catalogue access tool, CDS, Strasbourg, France \citep{ochsenbein_2000}. This work made use of \textsc{astropy}, a community-developed core \textsc{python} package and an ecosystem of tools and resources for astronomy (Astropy Collaboration \citep{astropy:2013}, \citep{astropy:2018}, \citep{astropy:2022}).



\bibliographystyle{mnras}
\bibliography{My_Library} 

\begin{thebibliography}{}
\makeatletter
\relax
\def\mn@urlcharsother{\let\do\@makeother \do\$\do\&\do\#\do\^\do\_\do\%\do\~}
\def\mn@doi{\begingroup\mn@urlcharsother \@ifnextchar [ {\mn@doi@} {\mn@doi@[]}}
\def\mn@doi@[#1]#2{\def\@tempa{#1}\ifx\@tempa\@empty \href {http://dx.doi.org/#2} {doi:#2}\else \href {http://dx.doi.org/#2} {#1}\fi \endgroup}
\def\mn@eprint#1#2{\mn@eprint@#1:#2::\@nil}
\def\mn@eprint@arXiv#1{\href {http://arxiv.org/abs/#1} {{\tt arXiv:#1}}}
\def\mn@eprint@dblp#1{\href {http://dblp.uni-trier.de/rec/bibtex/#1.xml} {dblp:#1}}
\def\mn@eprint@#1:#2:#3:#4\@nil{\def\@tempa {#1}\def\@tempb {#2}\def\@tempc {#3}\ifx \@tempc \@empty \let \@tempc \@tempb \let \@tempb \@tempa \fi \ifx \@tempb \@empty \def\@tempb {arXiv}\fi \@ifundefined {mn@eprint@\@tempb}{\@tempb:\@tempc}{\expandafter \expandafter \csname mn@eprint@\@tempb\endcsname \expandafter{\@tempc}}}

\bibitem[\protect\citeauthoryear{{Astropy Collaboration} et~al.,}{{Astropy Collaboration} et~al.}{2013}]{astropy:2013}
{Astropy Collaboration} et~al., 2013, \mn@doi [\aap] {10.1051/0004-6361/201322068}, \href {http://adsabs.harvard.edu/abs/2013A%26A...558A..33A} {558, A33}

\bibitem[\protect\citeauthoryear{{Astropy Collaboration} et~al.,}{{Astropy Collaboration} et~al.}{2018}]{astropy:2018}
{Astropy Collaboration} et~al., 2018, \mn@doi [\aj] {10.3847/1538-3881/aabc4f}, \href {https://ui.adsabs.harvard.edu/abs/2018AJ....156..123A} {156, 123}

\bibitem[\protect\citeauthoryear{{Astropy Collaboration} et~al.,}{{Astropy Collaboration} et~al.}{2022}]{astropy:2022}
{Astropy Collaboration} et~al., 2022, \mn@doi [\apj] {10.3847/1538-4357/ac7c74}, \href {https://ui.adsabs.harvard.edu/abs/2022ApJ...935..167A} {935, 167}

\bibitem[\protect\citeauthoryear{Bailer-Jones, Rybizki, Fouesneau, Demleitner  \& Andrae}{Bailer-Jones et~al.}{2021}]{bailer-jones_estimating_2021}
Bailer-Jones C. A.~L.,  Rybizki J.,  Fouesneau M.,  Demleitner M.,   Andrae R.,  2021, \mn@doi [The Astronomical Journal] {10.3847/1538-3881/abd806}, 161, 147

\bibitem[\protect\citeauthoryear{Brzycki, Siemion, de Pater, Cordes, Gajjar, Lacki  \& Sheikh}{Brzycki et~al.}{2023}]{brzycki_detecting_2023}
Brzycki B.,  Siemion A. P.~V.,  de Pater I.,  Cordes J.~M.,  Gajjar V.,  Lacki B.,   Sheikh S.,  2023, \mn@doi [The Astrophysical Journal] {10.3847/1538-4357/acdee0}, 952, 46

\bibitem[\protect\citeauthoryear{Choza, Croft, Siemion, Sheikh, Lebofsky, MacMahon, Drew  \& Worden}{Choza et~al.}{2024}]{choza_radio_2024}
Choza C.,  Croft S.,  Siemion A. P.~V.,  Sheikh S.,  Lebofsky M.,  MacMahon D. H.~E.,  Drew J.,   Worden S.~P.,  2024, \mn@doi [Research Notes of the American Astronomical Society] {10.3847/2515-5172/ad235f}, 8, 37

\bibitem[\protect\citeauthoryear{Cocconi \& Morrison}{Cocconi \& Morrison}{1959}]{cocconi_searching_1959}
Cocconi G.,  Morrison P.,  1959, \mn@doi [Nature] {10.1038/184844a0}, 184, 844

\bibitem[\protect\citeauthoryear{{Cordes} \& {Lazio}}{{Cordes} \& {Lazio}}{1993}]{Cordes_1993}
{Cordes} J.,  {Lazio} J.,  1993, in {Shostak} G.~S.,  ed.,  Astronomical Society of the Pacific Conference Series Vol. 47, Third Decennial US-USSR Conference on SETI. p.~143

\bibitem[\protect\citeauthoryear{Croft, Siemion, Cordes, Morrison, Paragi  \& Tarter}{Croft et~al.}{2018}]{croft_science_2018}
Croft S.,  Siemion A. P.~V.,  Cordes J.~M.,  Morrison I.~S.,  Paragi Z.,   Tarter J.,  2018, Science with an {ngVLA}: {SETI} {Searches} for {Evidence} of {Intelligent} {Life} in the {Galaxy}, \mn@doi{10.48550/arXiv.1810.06568}, \url {http://arxiv.org/abs/1810.06568}

\bibitem[\protect\citeauthoryear{Cullers}{Cullers}{1986}]{cullers_sensitive_1986}
Cullers D.~K.,  1986, \mn@doi [Acta Astronautica] {10.1016/0094-5765(86)90005-6}, 13, 31

\bibitem[\protect\citeauthoryear{Czech et~al.,}{Czech et~al.}{2021}]{czech_breakthrough_2021}
Czech D.,  et~al., 2021, \mn@doi [Publications of the Astronomical Society of the Pacific] {10.1088/1538-3873/abf329}, 133, 064502

\bibitem[\protect\citeauthoryear{Enriquez et~al.,}{Enriquez et~al.}{2017}]{enriquez_breakthrough_2017}
Enriquez J.~E.,  et~al., 2017, \mn@doi [The Astrophysical Journal] {10.3847/1538-4357/aa8d1b}, 849, 104

\bibitem[\protect\citeauthoryear{{Gaia Collaboration} et~al.,}{{Gaia Collaboration} et~al.}{2016}]{gaia_collaboration_gaia_2016}
{Gaia Collaboration} et~al., 2016, \mn@doi [Astronomy and Astrophysics] {10.1051/0004-6361/201629272}, 595, A1

\bibitem[\protect\citeauthoryear{{Gaia Collaboration} et~al.,}{{Gaia Collaboration} et~al.}{2023}]{gaia_collaboration_gaia_2023}
{Gaia Collaboration} et~al., 2023, \mn@doi [Astronomy and Astrophysics] {10.1051/0004-6361/202243940}, 674, A1

\bibitem[\protect\citeauthoryear{Gajjar et~al.,}{Gajjar et~al.}{2022}]{gajjar_searching_2022}
Gajjar V.,  et~al., 2022, \mn@doi [The Astrophysical Journal] {10.3847/1538-4357/ac6dd5}, 932, 81

\bibitem[\protect\citeauthoryear{Garrett}{Garrett}{2018}]{garrett_seti_2018}
Garrett M.~A.,  2018, {SETI} surveys of the nearby and distant universe employing wide-field radio interferometry techniques, \mn@doi{10.48550/arXiv.1810.07235}, \url {http://arxiv.org/abs/1810.07235}

\bibitem[\protect\citeauthoryear{Garrett \& Siemion}{Garrett \& Siemion}{2023}]{garrett_constraints_2023}
Garrett M.~A.,  Siemion A. P.~V.,  2023, \mn@doi [Monthly Notices of the Royal Astronomical Society] {10.1093/mnras/stac2607}, 519, 4581

\bibitem[\protect\citeauthoryear{{Horowitz} \& {Sagan}}{{Horowitz} \& {Sagan}}{1993}]{horowitz_meta}
{Horowitz} P.,  {Sagan} C.,  1993, \mn@doi [\apj] {10.1086/173157}, \href {https://ui.adsabs.harvard.edu/abs/1993ApJ...415..218H} {415, 218}

\bibitem[\protect\citeauthoryear{Kardashev}{Kardashev}{1964}]{kardashev_transmission_1964}
Kardashev N.~S.,  1964, Soviet Astronomy, 8, 217

\bibitem[\protect\citeauthoryear{{Leigh} \& {Horowitz}}{{Leigh} \& {Horowitz}}{2000}]{leigh_beta}
{Leigh} D.,  {Horowitz} P.,  2000, in {Lemarchand} G.,  {Meech} K.,  eds,  Astronomical Society of the Pacific Conference Series Vol. 213, Bioastronomy 99. p.~459

\bibitem[\protect\citeauthoryear{Mauersberger, Wilson, Rood, Bania, Hein  \& Linhart}{Mauersberger et~al.}{1996}]{mauersberger_seti_1996}
Mauersberger R.,  Wilson T.~L.,  Rood R.~T.,  Bania T.~M.,  Hein H.,   Linhart A.,  1996, Astronomy and Astrophysics, 306, 141

\bibitem[\protect\citeauthoryear{McGuire}{McGuire}{2022}]{mcguire_2021_2022}
McGuire B.~A.,  2022, \mn@doi [The Astrophysical Journal Supplement Series] {10.3847/1538-4365/ac2a48}, 259, 30

\bibitem[\protect\citeauthoryear{Ochsenbein}{Ochsenbein}{2000}]{ochsenbein_2000}
Ochsenbein F. e.~a.,  2000, { The VizieR database of astronomical catalogues }, \mn@doi{10.26093/cds/vizier}

\bibitem[\protect\citeauthoryear{Oliver}{Oliver}{1979}]{oliver_rationale_1979}
Oliver B.~M.,  1979, in Billingham J.,  Pešek R.,  eds, , Communication with {Extraterrestrial} {Intelligence}.
Pergamon, pp 71--79, \mn@doi{10.1016/B978-0-08-024727-4.50011-1}, \url {https://www.sciencedirect.com/science/article/pii/B9780080247274500111}

\bibitem[\protect\citeauthoryear{Oliver \& Billingham}{Oliver \& Billingham}{1971}]{oliver_project_1971}
Oliver B.~M.,  Billingham J.,  1971, The 1971 NASA/ASEE Summer Fac. Fellowship Program (NASA-CR-114445

\bibitem[\protect\citeauthoryear{Price et~al.,}{Price et~al.}{2018}]{price_breakthrough_2018}
Price D.~C.,  et~al., 2018, \mn@doi [Publications of the Astronomical Society of Australia] {10.1017/pasa.2018.36}, 35, e041

\bibitem[\protect\citeauthoryear{Price et~al.,}{Price et~al.}{2020}]{price_breakthrough_2020}
Price D.~C.,  et~al., 2020, \mn@doi [The Astronomical Journal] {10.3847/1538-3881/ab65f1}, 159, 86

\bibitem[\protect\citeauthoryear{{Rampadarath}, {Morgan}, {Tingay}  \& {Trott}}{{Rampadarath} et~al.}{2012}]{rampadarth_vlbi}
{Rampadarath} H.,  {Morgan} J.~S.,  {Tingay} S.~J.,   {Trott} C.~M.,  2012, \mn@doi [\aj] {10.1088/0004-6256/144/2/38}, \href {https://ui.adsabs.harvard.edu/abs/2012AJ....144...38R} {144, 38}

\bibitem[\protect\citeauthoryear{Remijan \& Markwick-Kemper}{Remijan \& Markwick-Kemper}{2007}]{splatalogue}
Remijan A.,  Markwick-Kemper A.,  2007, Bulletin of the American Astronomical Society, 39, 963

\bibitem[\protect\citeauthoryear{Sheikh, Wright, Siemion  \& Enriquez}{Sheikh et~al.}{2019}]{sheikh_choosing_2019}
Sheikh S.~Z.,  Wright J.~T.,  Siemion A.~P.,   Enriquez J.~E.,  2019, \mn@doi [The Astrophysical Journal] {10.3847/1538-4357/ab3fa8}, 884, 14

\bibitem[\protect\citeauthoryear{Sheikh, Siemion, Enriquez, Price, Isaacson, Lebofsky, Gajjar  \& Kalas}{Sheikh et~al.}{2020}]{sheikh_breakthrough_2020}
Sheikh S.~Z.,  Siemion A.,  Enriquez J.~E.,  Price D.~C.,  Isaacson H.,  Lebofsky M.,  Gajjar V.,   Kalas P.,  2020, \mn@doi [The Astronomical Journal] {10.3847/1538-3881/ab9361}, 160, 29

\bibitem[\protect\citeauthoryear{{Shirai}, {Oyama}, {Imai}  \& {Abe}}{{Shirai} et~al.}{2004}]{shirai_search_2004}
{Shirai} T.,  {Oyama} T.,  {Imai} H.,   {Abe} S.,  2004, in {Norris} R.,  {Stootman} F.,  eds,  IAU Symposium Vol. 213, Bioastronomy 2002: Life Among the Stars. p.~423

\bibitem[\protect\citeauthoryear{Siemion et~al.,}{Siemion et~al.}{2014}]{siemion_searching_2014}
Siemion A. P.~V.,  et~al., 2014, Searching for {Extraterrestrial} {Intelligence} with the {Square} {Kilometre} {Array}, \url {http://arxiv.org/abs/1412.4867}

\bibitem[\protect\citeauthoryear{{Steffes}}{{Steffes}}{1993}]{steffes_potential_1993}
{Steffes} P.,  1993, in {Shostak} G.~S.,  ed.,  Astronomical Society of the Pacific Conference Series Vol. 47, Third Decennial US-USSR Conference on SETI. p.~367

\bibitem[\protect\citeauthoryear{Steffes \& Deboer}{Steffes \& Deboer}{1993}]{steffes_deboer}
Steffes P.,  Deboer D.,  1993, Icarus, 107, 215

\bibitem[\protect\citeauthoryear{Townes}{Townes}{1983}]{townes}
Townes C.~H.,  1983, \mn@doi [Proceedings of the National Academy of Sciences] {10.1073/pnas.80.4.1147}, 80, 1147

\bibitem[\protect\citeauthoryear{Tremblay et~al.,}{Tremblay et~al.}{2023}]{tremblay_cosmic_2023}
Tremblay C.~D.,  et~al., 2023, {COSMIC}: {An} {Ethernet}-based {Commensal}, {Multimode} {Digital} {Backend} on the {Karl} {G}. {Jansky} {Very} {Large} {Array} for the {Search} for {Extraterrestrial} {Intelligence}, \url {http://arxiv.org/abs/2310.09414}

\bibitem[\protect\citeauthoryear{Wandia et~al.,}{Wandia et~al.}{2023}]{wandia_interferometric_2023}
Wandia K.,  et~al., 2023, \mn@doi [Monthly Notices of the Royal Astronomical Society] {10.1093/mnras/stad1151}, 522, 3784

\bibitem[\protect\citeauthoryear{Wlodarczyk-Sroka, Garrett  \& Siemion}{Wlodarczyk-Sroka et~al.}{2020}]{wlodarczyk-sroka_extending_2020}
Wlodarczyk-Sroka B.~S.,  Garrett M.~A.,   Siemion A. P.~V.,  2020, \mn@doi [Monthly Notices of the Royal Astronomical Society] {10.1093/mnras/staa2672}, 498, 5720

\bibitem[\protect\citeauthoryear{Worden et~al.,}{Worden et~al.}{2017}]{worden_breakthrough_2017}
Worden S.~P.,  et~al., 2017, \mn@doi [Acta Astronautica] {10.1016/j.actaastro.2017.06.008}, 139, 98

\bibitem[\protect\citeauthoryear{{Wright} et~al.,}{{Wright} et~al.}{2018}]{optical_shelley}
{Wright} S.~A.,  et~al., 2018, in {Evans} C.~J.,  {Simard} L.,   {Takami} H.,  eds,  Society of Photo-Optical Instrumentation Engineers (SPIE) Conference Series Vol. 10702, Ground-based and Airborne Instrumentation for Astronomy VII.

\bibitem[\protect\citeauthoryear{{Wrobel}}{{Wrobel}}{1995}]{wrobel_vlbi_1995}
{Wrobel} J.~M.,  1995, in {Zensus} J.~A.,  {Diamond} P.~J.,   {Napier} P.~J.,  eds,  Astronomical Society of the Pacific Conference Series Vol. 82, Very Long Baseline Interferometry and the VLBA. p.~411, \url {https://ui.adsabs.harvard.edu/abs/1995ASPC...82..411W}

\bibitem[\protect\citeauthoryear{{Zuckerman}}{{Zuckerman}}{2022}]{zuckerman}
{Zuckerman} B.,  2022, \mn@doi [\mnras] {10.1093/mnras/stac1113}, 514, 227

\makeatother
\end{thebibliography}




\bsp	
\label{lastpage}
\end{document}